\documentclass[10pt,journal,letterpaper]{IEEEtran}
\newcommand{\dmin}{\delta_{\mathrm{min}}}
\newcommand{\nmos}{\texttt{nMOS}}
\newcommand{\pmos}{\texttt{pMOS}}
\newcommand{\NOR}{\texttt{NOR}}

\usepackage{textcomp}
\usepackage{stfloats}
\usepackage{hyperref}   

\usepackage[cmex10]{amsmath}
\usepackage{amssymb,amsmath,amsthm}
\newtheorem{theorem}{Theorem}

\usepackage{algorithmic}
\usepackage{algorithm}

\usepackage{etoolbox}
\AtBeginEnvironment{algorithm}{\scriptsize}
\AtBeginEnvironment{algorithmic}{\scriptsize}

\usepackage{tikz}
\usetikzlibrary {shapes.multipart}
\usetikzlibrary {circuits.logic.US} 

\usepackage{circuitikz}

\usepackage{url}
\usepackage{todonotes}
\usepackage[caption=false,font=footnotesize]{subfig}

\usepackage[capitalise]{cleveref}

\usepackage{balance}

\begin{document}
\title{A Symbolic Execution Framework for Symbolic Timing Analysis of Digital Integrated Circuits\thanks{This work was funded in whole by the Austrian Science Fund (FWF) \href{https://www.fwf.ac.at/en/research-radar/10.55776/ESP1705325}{10.55776/ESP1705325} (\href{https://ucrisportal.univie.ac.at/en/projects/symbolische-zeitanalyse-asynchroner-schaltungen/}{STAAC Project}).}}

\author{
\IEEEauthorblockN{Dennis Eigner\IEEEauthorrefmark{1}, Arman Ferdowsi\IEEEauthorrefmark{2}, Ulrich Schmid\IEEEauthorrefmark{1}}

\IEEEauthorblockA{\IEEEauthorrefmark{1}Embedded Computing Systems Group, TU Wien, Vienna, Austria}
\IEEEauthorblockA{\IEEEauthorrefmark{2}Faculty of Computer Science, University of Vienna, Vienna, Austria \\
e11808235@student.tuwien.ac.at, arman.ferdowsi@univie.ac.at, s@ecs.tuwien.ac.at}
}

\maketitle

\begin{abstract}
Simulation-based dynamic timing analysis of digital integrated circuits (DDTA) offers a faster alternative to traditional analog SPICE simulations. To achieve timing predictions that are reasonably competitive in terms of accuracy, however, DDTA mandates gate delay models that go beyond the standard pure or inertial delay models used in state-of-the-art tools. Recent advances in analytic gate delay models, which now also capture effects like drafting and multi-input switching, unlock new possibilities for timing analysis, which go way beyond simulation-based approaches towards an exhaustive exploration. In this paper, we present the cornerstones of a novel symbolic execution framework, which utilizes such analytic delay models for automatically computing symbolic delay expressions for all paths in a digital circuit, for some given ordering of the input transitions. To reduce combinatorial explosion, we introduce symbolic pruning methods that also enable path-sensitive, goal-driven reasoning about timing properties and analytic optimization of specific circuit paths.
\end{abstract}
\IEEEpeerreviewmaketitle

\section{Introduction}

Our modern digital society primarily relies on billions of \emph{very-large-scale integrated} (VLSI) circuits, which control the operation of essentially every device today, ranging from medical pacemakers to household appliances to cars to data centers to large-scale distributed computer networks. Since modern VLSI circuits consist of millions, if not billions, of transistors that need to operate in a well-synchronized fashion, the importance of design methods that can guarantee correct circuit behavior cannot be overstated. At the same time, economic pressure demands reasonably short development cycles.

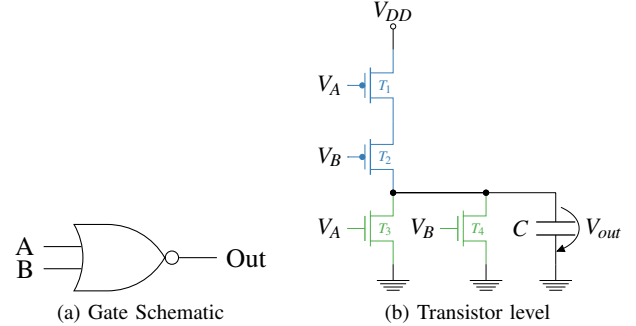
\begin{figure}[t]
  \centering
  \subfloat[Gate Schematic]{
\begin{tikzpicture}[scale=1,       transform shape, circuit logic US, huge circuit symbols]
    \node [nor gate] (a0) {};
    \draw (a0.input 1) -- ++(-0.5, 0) node[left] {A};
    \draw (a0.input 2) -- ++(-0.5, 0) node[left] {B};
    \draw (a0.output) -- ++(0.5, 0) node[right] {Out};
\end{tikzpicture}
    \label{fig:norgatedigital}}
  \hfil  
  \subfloat[Transistor level]{
\includegraphics[height=0.44\linewidth]{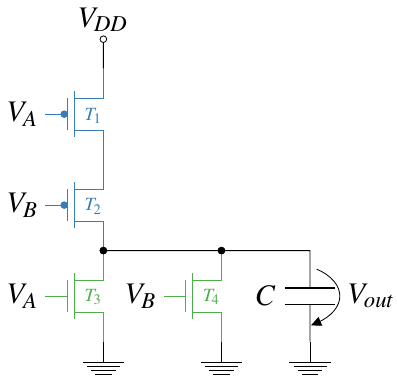}  
    \label{fig:norgateanalog}}
 
  \caption{\small Transistor schematic
of a CMOS NOR gate.}
\end{figure}

The key for accomplishing these conflicting goals is \emph{digital abstraction}: whereas the transistors that primarily form a VLSI circuit are inherently \emph{analog} electronic devices, which process information encoded in continuous-time, continuous-valued signals, they are essentially viewed as ideal switches that process and generate binary, event-based signals. Consider the schematic of a CMOS NOR gate shown in \cref{fig:norgateanalog}, for example, which consists of 4 transistors and one capacitor (representing the load caused by the successor gate(s)). Its inputs are represented by the time-dependent voltage values $V_A$ and $V_B$, and its output is $V_{out}$.

To obtain a digital abstraction for this NOR gate, one interprets an analog signal as HIGH if its voltage is above a certain threshold voltage, typically half the supply voltage, and LOW otherwise; an example can be found in \cref{fig:spice_simulation} and \cref{fig:digital_simulation}. As a consequence, the NOR gate can be described by a simple truth table. However, due to the finite signal propagation speed and, usually dominantly, the finite rise and fall times of analog signals, every gate causes a non-zero input-to-output delay. The core components of such a digital abstraction are hence \emph{gate delay models}, which ideally allow faithful modeling of the analog circuit behavior. Most state-of-the-art tools rely on \emph{static} delay models like pure or inertial delays \cite{Ung71}, which are typically parametrized using elaborate timing models like CCSM and ECSM \cite{Syn:CCSM,Cad:ECSM}. Despite considering corner cases only, static delay models facilitate \emph{static timing analysis} (STA) approaches, which are primarily used nowadays for validating the timing correctness of a digital circuit.

\subsection*{Static timing analysis}

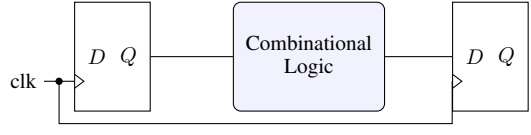
\begin{figure}[t]
    \centering
\begin{tikzpicture}[scale=0.8, transform shape]
\node[draw, minimum width=1.25cm, minimum height=1.8cm] (ff1) at (0,0) {};
\node[anchor=west] at ([xshift=1mm]ff1.west) {$D$};
\node[anchor=east] at ([xshift=-1mm]ff1.east) {$Q$};
\draw (ff1.west |- {$(ff1.south)+(0,0.38cm)$}) -- ++(0.16,0.12) -- ++(-0.16,0.12);

\node[draw, minimum width=2.5cm, minimum height=1.8cm,
      rounded corners=1mm, fill=blue!5, align=center]
      (comb) at (3.25,0) {Combinational\\ Logic};

\node[draw, minimum width=1.25cm, minimum height=1.8cm] (ff2) at (6.25,0) {};
\node[anchor=west] at ([xshift=1mm]ff2.west) {$D$};
\node[anchor=east] at ([xshift=-1mm]ff2.east) {$Q$};
\draw (ff2.west |- {$(ff2.south)+(0,0.38cm)$}) -- ++(0.16,0.12) -- ++(-0.16,0.12);

\draw (ff1.east) -- (comb.west);
\draw (comb.east) -- (ff2.west);

\coordinate (clk1) at (ff1.west |- {$(ff1.south)+(0,0.50cm)$});
\coordinate (clk2) at (ff2.west |- {$(ff2.south)+(0,0.50cm)$});
\draw (clk1) -- ++(-0.5,0) coordinate[pos=0.5] (clk_connection) node[left] {clk};
\draw (clk_connection) node[circle,fill,inner sep=1.2pt] {} -- ++(0,-0.7) -| (clk2);
\end{tikzpicture}
    \caption{\small Two stages of a synchronous circuit: flip-flops with combinatorial logic in between.}
    \label{fig:flipfop_setuphold}
\end{figure}

Consider a stage of the very common \emph{synchronous} circuit structure depicted in \cref{fig:flipfop_setuphold}. It consists of two flip-flops, driven by a common clock signal, with some combinatorial logic in between them. The purpose of the second flip-flop is to latch the current state of the output of the combinatorial logic when, say, a rising clock transition occurs and to hold it until the next clock transition. This way, the inputs of the combinatorial logic are kept stable for one clock period by the first flip-flop. 

\begin{figure}
    \centering

\begin{tikzpicture}[scale=0.8, transform shape, circuit logic US, huge circuit symbols]
\node [nor gate] (a0) {};
\node [nor gate, draw=red] (a1) at (0, -1.75) {};
\node [nor gate] (a2) at (0, -3.5) {};

\draw (a0.input 1) -- ++(-0.5, 0);
\draw (a0.input 2) -- ++(-0.5, 0) coordinate[pos=0.5] (a0_connection) {};

\draw[style={draw=red}] (a1.input 1) -- ++(-0.5, 0);
\draw[style={draw=red}] (a1.input 2) -- ++(-0.5, 0);

\draw (a2.input 1) -- ++(-0.5, 0);
\draw (a2.input 2) -- ++(-0.5, 0);

\draw (a1.output) coordinate (b0_anchor) {};
\node[nor gate, anchor=input 2] (b0) at (2,0 |- b0_anchor) {};

\draw (a0_connection) node[circ]{} -- ++(0, -0.5) -- ++(2, 0) |- (b0.input 1);
\draw (a1.output) -- (b0.input 2) coordinate[pos=0.5] (b1_connection) {};

\draw[style={draw=red}] (b1_connection) node[circ]{} -- ++(0, -1) coordinate (b1_anchor) {};
\node [not gate, anchor = input, draw=red] (b1) at (2,0 |- b1_anchor) {};

\draw[style={draw=red}] (a1.output) -- (b1_connection);
\draw[style={draw=red}] (b1_anchor) -- (b1.input);

\draw (a0.output) coordinate (c0_anchor) {};
\node [nor gate, anchor = input 1, draw=red] (c0) at (4,0 |- c0_anchor) {};

\draw (a0.output) -- (c0.input 1);
\draw[style={draw=red}] (b1.output) -- ++(0.5, 0) |- (c0.input 2);

\draw (a2.output) coordinate (c1_anchor) {};
\node [nor gate, anchor = input 2] (c1) at (4,0 |- c1_anchor) {};

\draw (b0.output) -- ++(0.25, 0) |- (c1.input 1);
\draw (a2.output) -- (c1.input 2);

\draw[style={draw=red}] (c0.output) -- ++(0.5, 0) node[right, draw=red] {Worst-case};
\draw (c1.output) -- ++(0.5, 0);
\end{tikzpicture}
    
    \caption{\small Possible worst-case data path of a combinatorial logic (caused by an
    inverter that is slow compared to the $\NOR$ gates).}
    \label{fig:worst_case}
\end{figure}
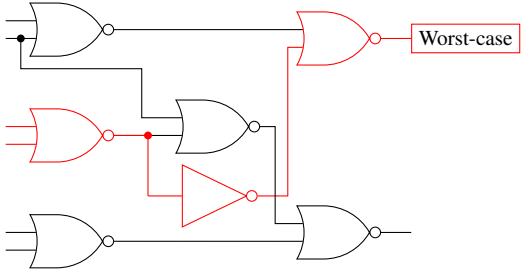

\begin{figure}
    \centering
    \includegraphics[width=0.48\textwidth]{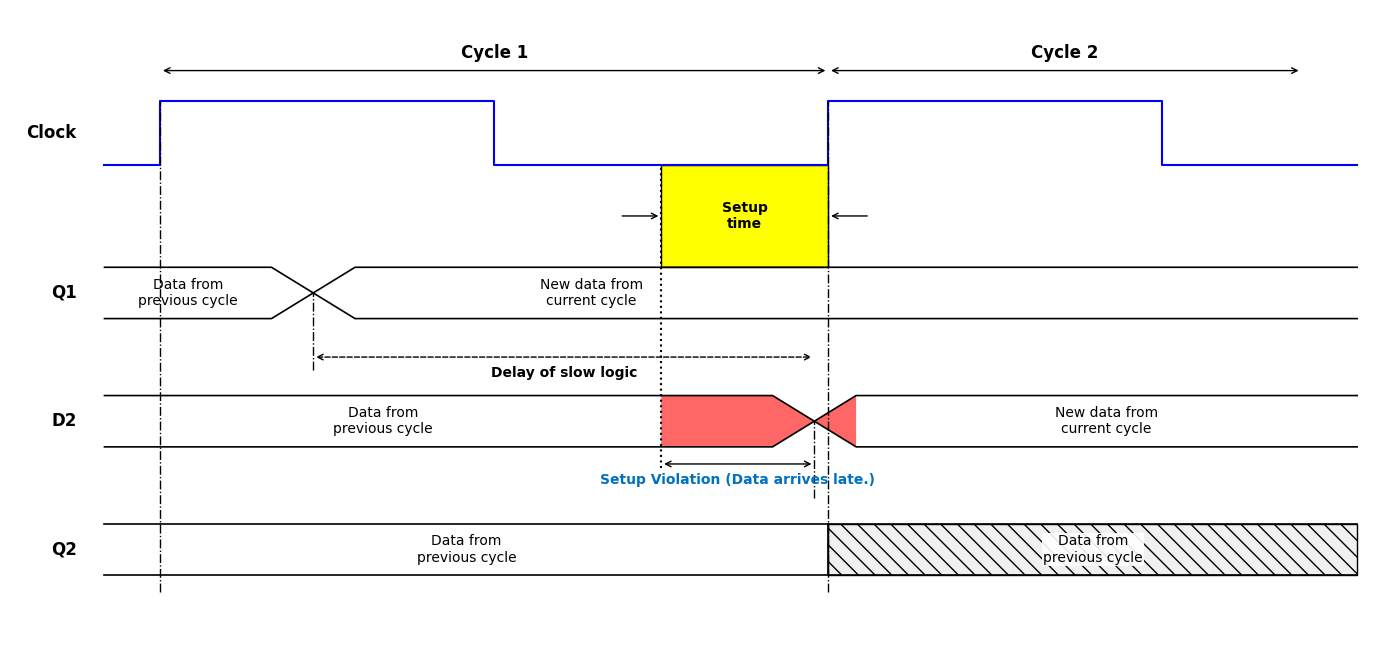}
    \caption{\small Example timing diagram for a setup-hold time violation in a synchronous circuit as depicted in \cref{fig:flipfop_setuphold}. $Q1$ represents the output of the
    first flip-flop, $D2$ (resp.\ $Q2$) the data input (resp.\ the output) of the second flip-flop.}
    \label{fig:setup_hold}
\end{figure}

To ensure correct circuit behavior, however, certain timing constraints must be enforced. In particular, the \emph{setup} and \emph{hold} times of the flip-flops must be obeyed; see \cref{fig:setup_hold} for details: The output of the combinatorial logic must be stable already some time before a clock transition occurs and must remain stable for some time after it as well. Static timing analysis approaches are sufficient for validating this, as it suffices to determine the delay of the worst-case path(s) in the combinatorial logic (see \cref{fig:worst_case}) and compare it to the clock period. And indeed, modern simulation-based \emph{statistical} static timing analysis approaches can validate even very large circuit designs in reasonable time nowadays.

However, traditional static timing analysis approaches also have some deficiencies: (i) they might provide false positives of errors, due to focusing on worst-case delays only, and (ii) they lack the ability to identify the root causes of a timing violation. Both limit the utility of static timing analysis from a design-optimization viewpoint.

\subsection*{Dynamic timing analysis}

\begin{figure}[h]
  \centering
  \subfloat[Analog simulation trace]{
\includegraphics[height=0.18\linewidth]{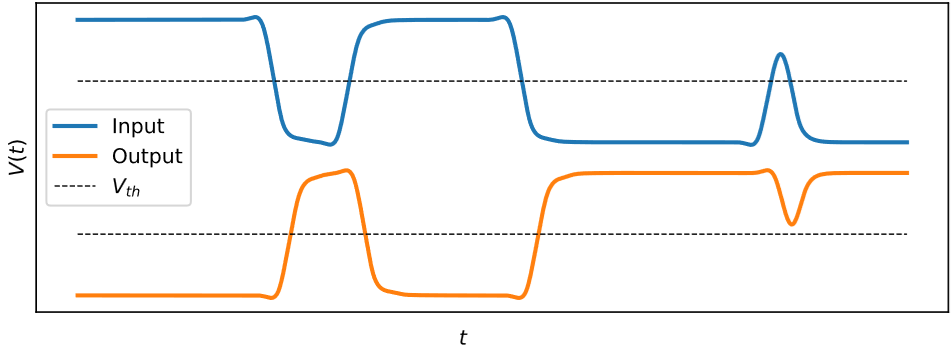}  
    \label{fig:spice_simulation}}
  \hfil
  \subfloat[Corresponding digital trace]{
 \includegraphics[height=0.18\linewidth]{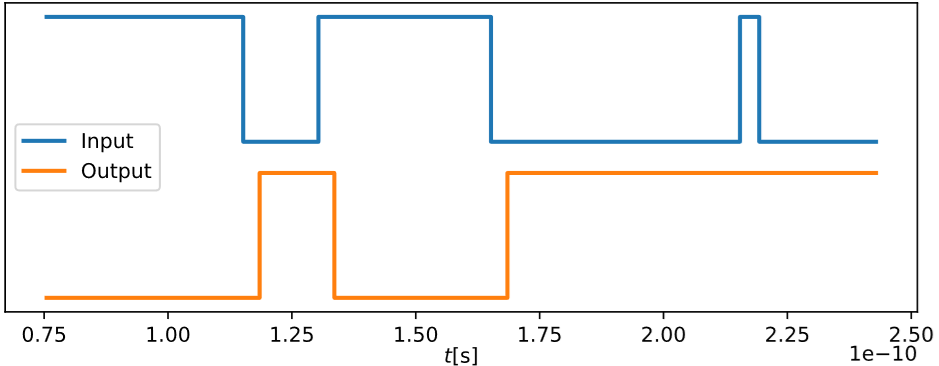}%
    \label{fig:digital_simulation}}    
  \caption{\small Example of an analog simulation trace and a digital one.}
\end{figure}

To explore both (i) and (ii) systematically, one usually has to resort to full analog simulations, applied to a small core part of the circuit that causes the problem. 
Analog simulation tools such as SPICE \cite{NP73:spice} simulate the system of 
differential equations representing the transistors in the circuit (recall \cref{fig:norgateanalog}), which are provided by the manufacturer of the circuit implementation technology. An example of the result of such a simulation is shown in \cref{fig:spice_simulation}. The major disadvantage of this approach, however, is that it is extremely slow. In reality, it is
impossible to even simulate even a short signal trace of a just moderately large circuit, say, a few ten microseconds of a circuit consisting of a few hundred gates.

A much faster alternative are \emph{dynamic digital timing analysis} (DDTA) approaches, see e.g.\ \cite{FFSS26:Integration} for a short overview, which use discrete-event simulation based on gate delay models. The results of such a simulation run are shown in \cref{fig:digital_simulation}. However, state-of-the-art DDTA tools like Questa, which are based on static pure or inertial gate delay models, offer modest timing prediction accuracy only. This can be improved by using \emph{history-aware} delay models as introduced in \cite{BJV06,FNS16:ToC,FNNS19:TCAD}, where the delay for some given input transition also depends on previous transitions. 

The recent family of history-aware hybrid gate delay models for multi-input gates introduced in \cite{ferdowsi2025faithful,FFSS26:Integration,ferdowsi2025drafting} even provide analytic closed-form gate delay formulas, i.e., deterministic functions, which accurately cover both \emph{multi-input switching} (delay variations due to close transitions at different inputs) and even \emph{drafting effects} (delay variations due to succeeding transitions), as revealed in \cite{ferdowsi2025drafting}. Moreover, these models can even model PVT effects, as e.g.\ delay dependencies cause by supply voltage variations are easy to incorporate.
Employing such advanced gate delay models in DDTA substantially outperforms state-of-the-art DDTA tools in terms of accuracy.

\subsection*{Unlocking symbolic timing analysis}

Besides boosting the accuracy of simulation-based DDTA approaches, the analytic gate delay formulas provided in \cite{ferdowsi2025drafting} also unlock a fundamentally different approach for timing analysis: since these formulas (i) are \emph{deterministic functions} that only involve symbolic gate parameters and (ii) can be composed to \emph{accurately} compute the delay of a path made up by several interconnected gates, they are amenable to symbolic manipulation as well as mathematical sensitivity analysis. Consequently, they provide a promising basis for \emph{symbolic} timing analysis of digital circuits, which is the objective of this paper. In sharp contrast to any simulation-based approach, be it SPICE or DDTA, our approach provides an \emph{exhaustive} characterization of all possible traces generated by a circuit. The obvious downside is the resulting huge computational complexity, which is, however, relativized by (i) the availability of effecive complexity reduction techniques and (ii) the fact that one usually exhaustively explores only a small core part 
of a circuit anyway.

\medskip

\noindent
\textbf{Main contributions:}
We introduce the cornerstones of a novel symbolic execution framework facilitating symbolic timing analysis of digital integrated circuits: given a circuit, its initial state, and an ordered sequence of symbolic input transition times, it uses the gate delay models of the involved gates to compute symbolic occurrence-time expressions for the transitions reached along every explored path, together with the corresponding path constraints. The framework consists of two main components:

(1) A novel symbolic execution engine, which is used for building a structured, delay-model-agnostic representation of the symbolic state-space tree of the circuit. We stress already here that, compared to software programs, the state-space of a circuit is very simple, as it only comprises the vector of the outputs of every gate (plus the queue of transitions of the external inputs). This is a consequence of the fact that gates are combinatorial logic elements and thus do not have an internal state caused by variables etc. Nevertheless, gates
do have a \emph{delay-induced internal state}, which is not considered in standard symbolic execution: it is caused by the non-instantaneous reaction of a gate output on relevant gate input changes.

Consequently, for a given order of external input transitions, our custom symbolic execution engine has to compute a path for \emph{every} possible transition order that might occur in an execution and allows for compact representation of simple loops. Whereas complex loops must be unrolled in our approach, we also introduce the concept of a \emph{meta-transition}, which can be viewed as a symbolic transition that represents all loop iterations. 

(2) A mechanism for annotating the tree nodes with explicit symbolic timestamps (that only involve the symbolic input transition times and the gate parameters), which are computed from the analytic delay formulas provided by the gate delay models by a computer algebra system \cite{TEFS25:arxiv}. 

Note that we avoid constructing non-differentiable timing formulas involving $\min/\max$ operators, but rather stitch together the analytic formulas describing the gate delays along a given path. Moreover, we augment nodes with symbolic constraints that are accrued to path constraints via constraint propagation, which enables very effective pruning. Note that our approach does not need a restricted constraint
language, since we allow any constraint that can be specified via a mathematical 
formula involving symbolic timestamps. 

Our framework also supports the selection of the actual paths of interest via \emph{goal functions}. A suitably defined goal function allows to restrict the exploration to those paths that contain contain specific hazards (i.e., bad transition orderings), and thus allows to find conditions for avoiding those, or to determine the minimal and/or maximal delay between certain transitions. Both are relevant for analyzing the root cause of setup/hold time violations, for example. 

We also provide a glimpse of the performance achieved by a (non-engineered) research prototype implementation of our approach for the benchmarking circuit c17\_slack from the ISCAS85 benchmarking library \cite{ISCAS85_reference}. Whereas our results are, of course, by no means representative, they show that our symbolic pruning method is already very effective.

\noindent
\textbf{Paper organization:} \cref{sec:relatedwork} provides an overview of the (few) related papers on symbolic timing analysis of integrated circuits in the literature. \cref{sec:modelandassumptions} introduces the underlying circuit model and some terms used throughout the paper. \cref{sec:symbolicexecution} explains how to construct the basic state space tree, including loop handling, and provides a coarse analysis of the worst-case complexity of our method. \cref{sec:delayformulas} explains our tree augmentation with symbolic timestamps and constraints and the way constraint propagation is used for pruning to mitigate combinatorial explosion. \cref{sec:integration} provides a brief description of the core algorithms of our complete symbolic timing analysis framework, as well as the results of our preliminary benchmark experiments. \cref{sec:conclusions} concludes our paper. A glimpse of our current work on additional partial order reduction techniques is outlined in Appendix \ref{app:discussion}.

\section{Related work \label{sec:relatedwork}}

For an overview of the rich state of the art of traditional static and dynamic simulation-based digital timing analysis, the reader is referred to \cite{FFSS26:Integration}. A remarkable recent simulation-based digital timing analysis approach, which also supports design optimization to some extent, is NVIDIA's INSTA framework \cite{luinsta}. In essence, it replaces the $\min$ and $\max$ functions traditionally used for describing gate timing behavior by soft-$\min$ and soft-$\max$, and leverages gradient-descent methods from machine learning to optimize critical delay values.

Whereas there is a rich literature on general symbolic execution (see \cite{BCDD19:survey} for a survey), to the best of our knowledge, only very few early papers have tried to employ those techniques for symbolic timing analysis of integrated circuits. More specifically, we are only aware of the following truly related work:

Ishiura et al.\ \cite{ishiura1989time} introduced a technique termed Time-Symbolic Simulation for accurate timing verification of combinatorial logic circuits. It treats gates either as pure delay gates or as purely functional gates. An iterative approach is used to construct a tree of possible events given an input sequence. Downsides of their approach are the inability to handle feedback loops and the restriction to pure gate delay models. Like any symbolic execution-based technique, evaluating larger circuits becomes computationally expensive; the authors could handle circuits with up to 100 gates, however.

Ishiura et al.\ \cite{ishiura1991coded} followed up to remedy some of these issues in their follow-up work called Coded Time-Symbolic Simulation. Here, the authors went on to use a shared binary decision diagram to represent the possible events of the circuit. This substantially reduced the computational costs but did not resolve the other deficiencies. Maler et al.\ \cite{maler1995timing} showed how to convert combinatorial circuits into timed automata, on which reasoning is much easier. The authors noted, however, that this conversion is not trivial for general circuits and hence part of ongoing research. Building on a timed automata-based approach, Claris{\'o} et al.\ \cite{clariso2004verification} used convex polyhedra to find a set of constraints on delay parameters to guarantee correct system behavior. Operations on these polyhedra are, again, computationally expensive and not feasible for circuits with more than 15 gates.

\section{Model and Assumptions \label{sec:modelandassumptions}}

We will rely on the thresholded hybrid model of digital circuits and their executions, which we introduced in \cite[Sec.~4]{ferdowsi2025faithful}: We assume that circuits process and generate \emph{binary} signals, represented by finite or infinite sequences of \emph{transitions} $(v_i,t_i)$ with $v_i\in\{0,1\}$, strictly increasing occurrence times $t_i\in\mathbb{R}$ for ordinary transitions, and alternating values. A transition $(0,t)$ denotes a \emph{falling transition} at time $t$, and $(1,t)$ denotes a \emph{rising transition}. Note that \cite{ferdowsi2025faithful} assumes that the initial state of a circuit is determined by initial transitions $(0,-\infty)$ or $(1,-\infty)$, happening at time $-\infty$, and that the actual execution of a circuit starts at time $t=0$.

A gate is internally modeled by a \emph{thresholded hybrid automaton}. It is described by a system of first-order differential equations (with non-constant coefficients), where the mode switches are governed by the digital inputs, and the digital output is generated by digitizing the analog output signal using a threshold voltage comparator. Its behavior is described by a detailed digital \emph{gate delay model}, the formulas of which involve a number of \emph{gate parameters} that can be used to parametrize the model for a given real gate.

\emph{Circuits} are obtained by interconnecting a set of input ports and a set of output ports. The ports form the external interface of a circuit and a finite set of gates like NOR, NAND, etc., where a gate delay model like the ones in \cite{FFSS26:Integration} is available. We do not restrict how gates are interconnected in a circuit, except that we disallow connections between the output ports of two different gates and/or of the circuit. Formally, a {\em circuit\/} is described by a directed graph where:
\begin{enumerate}
\item[C1)] A vertex $\Gamma$ can be either a circuit input port, a circuit output port, or a digitized hybrid gate.
\item[C2)] An edge $(\Gamma,I,\Gamma')$ represents a zero-delay connection from
the output of $\Gamma$ to the fixed input $I$ of $\Gamma'$.
\item[C3)] Circuit input ports have no incoming edges.
\item[C4)] Circuit output ports have exactly one incoming edge and no outgoing one. 
\item[C5)] A $c$-ary gate $G$ has a single output and $c$ inputs $I_1,\dots, I_c$, in a fixed order, fed by incoming edges from exactly one gate output or one input port.
\end{enumerate}

\cite{ferdowsi2025faithful} also formally defines \emph{executions} of a circuit. An execution starts from a given sequence of transitions at every circuit input port, collectively termed as \emph{circuit input transitions}. Under the assumption that all gates in a circuit are \emph{strictly causal} (see \cite[Def.~2]{ferdowsi2025faithful}), in the sense that their input-to-output delays are strictly positive, one can prove that executions are unique (albeit the do of course depend on the choice of the gate parameters). Note carefully that this uniqueness of executions is instrumental for the feasibility of our approach, which aims at precisely characterizing the symbolic timing behavior of all possible executions of a circuit.

\begin{theorem}[Unique execution {\cite[Thm.~4.1]{ferdowsi2025faithful}}]\label{thm:execution}
Every circuit $C$ made up of finitely many strictly causal thresholded hybrid
gates has a unique execution, which either consists of finitely
many transitions only or else requires $[0,\infty)$ as its time
domain.
\end{theorem}

Rather than dealing with the point-to-point edges defined in C2), we will subsequently focus on \emph{wires} to describe the interconnect of a circuit. A wire just subsumes the set of edges that start either in a single circuit input port or in a single gate output. Consequently, a single wire compactly describes the common situation of a fan-out of an output that drives multiple input ports. The set of all wires will be denoted by $W=W_I \cup W_G$, where $W_I$ and $W_G$ denote the wires that start in a circuit input port and a gate output port, respectively. Note that the number of wires $|W|$ in a circuit is equal to the number of gates $|W_G|$ plus the number of circuit input ports $|W_I|$.

\section{Symbolic Execution for Timing Analysis}
\label{sec:symbolicexecution}

The core idea of our symbolic timing analysis approach is to use formulas, not time values, for expressing circuit timing behavior in terms of an ordered sequence of symbolic input transition times and symbolic gate parameters. The latter are supplied via the analytic delay formulas provided by gate delay models such as \cite{FFSS26:Integration,ferdowsi2025drafting}, which can be used to compute symbolic delay expressions via a computer algebra system such as SageMath or MATLAB; see \cite{TEFS25:arxiv} for some details. For example, the following expression is a representative piecewise delay formula for a two-input \NOR\ gate and a fixed local history.
\begin{align*}
\Delta_*(T)
  &=\frac{(\alpha_1+\alpha_2)
      (\delta_0(T)-\delta_\infty(T))}{\alpha_1},\\
\eta(T)
  &=-2RC_3\log\!\left(
     \frac{1}{2-\exp\!\left(
       -\frac{T+\dmin}{C'_1R_{n_B}}
     \right)}\right)+\dmin.
\end{align*}
\begin{equation*}
\delta^{\uparrow}(\Delta,T)\approx
\begin{cases}
\delta_0(T)-\dfrac{\alpha_1}{\alpha_1+\alpha_2}\Delta+\dmin,
  & \mathcal R_1,\\[1mm]
\delta_\infty(T)+\dmin,
  & \mathcal R_2,\\[1mm]
\eta(T),
  & \mathcal R_3,
\end{cases}
\end{equation*}
where
\begin{align*}
\mathcal R_1&:\ T+\dmin\geq0,\quad
                  0\leq\Delta<\Delta_*(T),\\
\mathcal R_2&:\ T+\dmin\geq0,\quad
                  \Delta\geq\Delta_*(T),\\
\mathcal R_3&:\ T+\dmin<0.
\end{align*}
Here $\Delta=t_B-t_A\geq 0$. The symbols $t_A$ and $t_B=t$ denote the last falling input transition time of input $A$ and input $B$, respectively, before or at $t$ (or $-\infty$ if none).
$\alpha_1,\alpha_2$ and $R$ are resistance parameters of the serial \pmos\ transistors in the \NOR\ gate. The symbol $d_{\min}$ denotes a pure delay, $C_3$ is the effective load capacitance, $C'_1$ and $R_{n_B}$ are, respectively, the capacitance and \nmos-resistance parameters appearing in the exponential branch. In \cite{FFSS26:Integration}, we showed that the fitted transistor-resistance parameters can be computed analytically from six characteristic
gate delay values (three for rising and falling transitions each, for $\Delta=0$, $\Delta=\infty$ and
$\Delta=-\infty$).

Note carefully that our symbolic timing analysis 
approach rests on the fact that gate delays are
deterministic functions of the gate parameters, possible including PVT-related ones like the
supply voltage, i.e., do not involve any statistical or non-deterministic uncertainties.
We consider this as an advantage of our approach, as it avoids the inevitable blow-up of
the uncertainties when composing gates in a path.

At the core of our approach is a symbolic execution engine. It constructs a \emph{state-space tree} that enumerates all possible states a circuit may take during any possible execution that starts out from a \emph{fixed} order of the sequence of symbolic input transition times. Informally, every \emph{node} in the tree encodes a particular \emph{state} of the circuit, which is just an assignment of binary values to all wires of the circuit. Later, every node will also be annotated with a set of timing conditions (\emph{constraints}) that must hold in an execution to reach the given state.

The successors of a node are determined by \emph{single} transitions that may occur in the corresponding state, forming the \emph{edges} in the tree. An edge corresponds either to (i) a single input transition or (ii) to a single output transition of an \emph{inconsistent gate}. The latter is characterized by the property that the value of the gate output does not match the one corresponding to the gate inputs in the current state; such a state is called an \emph{inconsistent state}. For example, consider a circuit state where a NOR gate has 0 applied to both inputs and therefore outputs a 1. If one of the inputs experiences a rising transition, the inputs are now 1 and 0, which would demand the output to be 0. Since the gate's output remains 1 some time after that single transition, however, it is inconsistent. Our construction guarantees that a \emph{path} in the state-space tree corresponds to a specific order of the (symbolic) transitions occurring in the circuit.

Note that we deliberately restrict our attention to executions for a given order of the input transitions here, as this is the most common situation in dynamic timing analysis. This restriction also dramatically reduces the number of executions of a circuit. It should be noted, though, that it is of course also possible to extend our approach to arbitrary \emph{sets} of different orderings.

\subsection{Basic state-space tree construction}
\label{sec:constructingtree}
Formally, our state-space tree is a directed tree $T = (S, E)$ (which will be turned into
a tree-like directed acyclic graphs by meta-transitions introduced below), where
\begin{itemize}
  \item A \emph{state}, $s \in S$, is a node in the tree representing the state of the entire circuit at some point in an execution. It is given as an assignment of binary values to all wires $w \in W$ and also encompasses the (remaining) \emph{input queue} $Q$, containing all symbolic input transitions that have not been processed yet. Note that we will augment the circuit state with a symbolic timestamp and additional delay constraints later on; see \cref{sec:delayformulas}.
  \item A \emph{transition}, $e \in E$, is an edge between two states ($s_i, s_j$), representing a transition of a single wire $w \in W$. We denote a rising (resp.\ falling) transition of $w$ as $w^+$ (resp.\ $w^-$). Note that our model does not consider concurrent transitions but rather requires an explicit ordering of any two transitions that happen at the same time.
\end{itemize}

The tree construction algorithm for a circuit $C$ will start out from a single \emph{root node}, which represents the initial state $s_0$ of the circuit.\footnote{In \cite{ferdowsi2025faithful}, the initial state of a circuit is actually determined by initial input transitions happening at time $t=-\infty$; recall \cref{sec:modelandassumptions}. For simplicity, we will assume here that the initial state is explicitly given, however.} Its evolution is determined by a given sequence of symbolic input transitions (starting at or after $t=0$) given in the initial \emph{input queue} $Q$ of finite size $|Q|$. Note that input transitions are read and removed from left to right (in the timing order) from $Q$ during tree construction. We write $Q=\bigl((w_1,t_1),\ldots,(w_m,t_m)\bigr)$ and assume that it is a valid input sequence: the symbolic occurrence times satisfy $0 \leq t_1 \leq t_2 \leq \cdots \leq t_m$, and $w_i \in \{w^-, w^+\}$ for some
input wire $w$. Moreover, for each input wire $w$, successive input transitions must have strictly increasing occurrence times and alternate in value. Note that these input-order constraints will be part of the root path constraints added in \cref{sec:delayformulas}.

\cref{alg:combine} tracks the current state $SC$ of the circuit during tree construction and uses two subroutines: \cref{alg:construct} for building a new node in the tree and \cref{alg:enumerate} for determining the successors of a node. First, it constructs the root node (variable $rootNode$) corresponding to the initial state $SC=s_0$ of the circuit and the initial input queue $Q$. Starting from $rootNode$, it then uses a local \emph{node queue} (variable $queue$) to organize a standard breadth-first construction of the tree. 

\begin{algorithm}
\caption{\small Construct full tree}
\label{alg:combine}
\begin{algorithmic}
\REQUIRE Circuit $C$, initial state $s_0$, initial input queue $Q$
\ENSURE Full tree constructed
\STATE $rootNode \leftarrow$ constructNode($s_0$, $Q$)
\STATE $queue \leftarrow \emptyset$
\STATE $queue.$append($rootNode$)
\WHILE{$queue.$length $> 0$}
    \STATE $nextItem \leftarrow queue.$popleft()
    \STATE createChildren($nextItem.SC$, $nextItem.Q$, $nextItem$)
    \FORALL{$child$ in $nextItem.children$}
        \STATE $queue$.append($child$)
    \ENDFOR
\ENDWHILE
\end{algorithmic}
\end{algorithm}

Tree nodes are constructed using \cref{alg:construct}. The procedure takes as input the current circuit state $SC$ and the current input queue $Q$. It uses the subroutine calculateInconsistencies($SC$) for computing the newly generated node $N$'s inconsistency queue $N.IQ$, which contains a list of output transitions of the inconsistent gates in the circuit state $SC$ (if any). Note carefully that the latter is calculated from the circuit state parameter $SC$ and not just passed down from $N$'s parent. Once a node is constructed, all its entries except its children are immutable.

\begin{algorithm}
\caption{\small Procedure constructNode}
\label{alg:construct}
\begin{algorithmic}
\REQUIRE Circuit state $SC$, input queue $Q$
\ENSURE A valid tree node $N$
\STATE $N \leftarrow$ newNode()
\STATE $N.SC \leftarrow SC$
\STATE $N.Q \leftarrow Q$
\STATE $N.IQ \leftarrow$ calculateInconsistencies($SC$)
\STATE $N.children \leftarrow \emptyset$
\RETURN $N$
\end{algorithmic}
\end{algorithm}

The core of the tree construction algorithm is formed by the procedure createChildren given in \cref{alg:enumerate}, which adds the set of child nodes to a previously constructed node $N$; it gets $N$'s circuit state $SC$, its queue $Q$, and $N$ itself as parameters. If $Q \neq \emptyset$, it first constructs a child node corresponding to the first input transition (variable $inputTransition$) in the input queue $Q$; the child node inherits $Q$ after popping $inputTransition$, as well as the circuit state resulting from applying $inputTransition$ to $SC$. Subsequently, the procedure iterates over all inconsistency transitions (variable $inconsistencyTransition$) in $N$'s inconsistency queue $N.IQ$, where it creates and adds a corresponding child for each of those. Note that the resulting child nodes inherit the input queue $Q$ unchanged, and the circuit state resulting from applying $inconsistencyTransition$ to $SC$. (Recall that each transition, regardless of whether it comes from the input queue or from the inconsistency queue, can only change the state of a single wire.)

\begin{algorithm}
\caption{\small Procedure createChildren}
\label{alg:enumerate}
\begin{algorithmic}
\REQUIRE Circuit state $SC$, input queue $Q$, tree node $N$
\ENSURE All valid children added to $N$
\IF{$Q.\mathrm{length} > 0$}
    \STATE $newQueue \leftarrow Q.\mathrm{copy}()$
    \STATE $inputTransition \leftarrow newQueue.\mathrm{popleft}()$
    \STATE $newCircuitstate \leftarrow SC.\mathrm{copy}()$
    \STATE $newCircuitstate.\mathrm{applyTransition}(inputTransition)$
    \STATE $N.\mathrm{addChild}(\mathrm{constructNode}(newCircuitstate, newQueue))$
\ENDIF
\FORALL{$inconsistencyTransition$ in $N.IQ$}
    \STATE $newCircuitstate \leftarrow SC.\mathrm{copy}()$
    \STATE $newCircuitstate.\mathrm{applyTransition}(inconsistencyTransition)$
    \STATE $N.\mathrm{addChild}(\mathrm{constructNode}(newCircuitstate, Q))$
\ENDFOR
\end{algorithmic}
\end{algorithm}

\subsection{Examples}
\label{Sec_Example}
\begin{figure}[h]
  \centering
  \begin{tikzpicture} [ scale=0.69,       transform shape,
    baseline=0,
    level distance = 20mm,
    sibling distance = 30mm
]
\node [rectangle split, rectangle split parts=3, draw] {
    \textbf{\color{red}1001} 
    \nodepart{second} 
    \scriptsize Q: $A^-$, $B^+$
    \nodepart{third} 
    \scriptsize IQ: $\emptyset$
}
child {
    node[rectangle split, rectangle split parts=3, draw] {
        \textbf{\color{red}0001} 
        \nodepart{second} 
        \scriptsize Q: $B^+$
        \nodepart{third} 
        \scriptsize IQ: $C^+$
    }
    child {
        node[rectangle split, rectangle split parts=3, draw] {
            \textbf{\color{red}0101} 
            \nodepart{second} 
            \scriptsize Q: $\emptyset$
            \nodepart{third} 
            \scriptsize IQ: $D^-$
        }
        child {
            node[rectangle split, rectangle split parts=3, draw] {
                \textbf{\color{red}0100} 
                \nodepart{second} 
                \scriptsize Q: $\emptyset$
                \nodepart{third} 
                \scriptsize IQ: $\emptyset$
            }
            edge from parent node[midway, left] {$D^-$}
        }
        edge from parent node[midway, left] {$B^+$}
    }
    child {
        node[rectangle split, rectangle split parts=3, draw] {
            \textbf{\color{red}0011} 
            \nodepart{second} 
            \scriptsize Q: $B^+$
            \nodepart{third} 
            \scriptsize IQ: $D^-$
        }
        child {
            node[rectangle split, rectangle split parts=3, draw] {
                \textbf{\color{red}0111} 
                \nodepart{second} 
                \scriptsize Q: $\emptyset$
                \nodepart{third} 
                \scriptsize IQ: $C^-$, $D^-$
            }
            child {
                node[rectangle split, rectangle split parts=3, draw] {
                    \textbf{\color{red}0101} 
                    \nodepart{second}
                    \scriptsize Q: $\emptyset$
                    \nodepart{third} 
                    \scriptsize IQ: $D^-$
                }
                child {
                    node[rectangle split, rectangle split parts=3, draw] {
                        \textbf{\color{red}0100} 
                        \nodepart{second} 
                        \scriptsize Q: $\emptyset$
                        \nodepart{third} 
                        \scriptsize IQ: $\emptyset$
                    }
                    edge from parent node[midway, left] {$D^-$}
                }
                edge from parent node[midway, left] {$C^-$}
            }
            child {
                node[rectangle split, rectangle split parts=3, draw] {
                    \textbf{\color{red}0110} 
                    \nodepart{second} 
                    \scriptsize Q: $\emptyset$
                    \nodepart{third} 
                    \scriptsize IQ: $C^-$
                }
                child {
                    node[rectangle split, rectangle split parts=3, draw] {
                        \textbf{\color{red}0100} 
                        \nodepart{second} 
                        \scriptsize Q: $\emptyset$
                        \nodepart{third} 
                        \scriptsize IQ: $\emptyset$
                    }
                    edge from parent node[midway, left] {$C^-$}
                }
                edge from parent node[midway, left] {$D^-$}
            }
            edge from parent node[midway, left] {$B^+$}
        }
        child {
            node[rectangle split, rectangle split parts=3, draw] {
                \textbf{\color{red}0010} 
                \nodepart{second} 
                \scriptsize Q: $B^+$
                \nodepart{third} 
                \scriptsize IQ: $\emptyset$
            }
            child {
                node[rectangle split, rectangle split parts=3, draw] {
                    \textbf{\color{red}0110} 
                    \nodepart{second} 
                    \scriptsize Q: $\emptyset$
                    \nodepart{third} 
                    \scriptsize IQ: $C^-$
                }
                child {
                    node[rectangle split, rectangle split parts=3, draw] {
                        \textbf{\color{red}0100} 
                        \nodepart{second} 
                        \scriptsize Q: $\emptyset$
                        \nodepart{third} 
                        \scriptsize IQ: $\emptyset$
                    }
                    edge from parent node[midway, left] {$C^-$}
                }
                edge from parent node[midway, left] {$B^+$}
            }
            edge from parent node[midway, left] {$D^-$}
        }
        edge from parent node[midway, left] {$C^+$}
    }
    edge from parent node[midway, left] {$A^-$}
};
\end{tikzpicture}
  \caption{\small Example of a basic state-space tree without loops.}
  \label{fig:full_tree}
\end{figure}
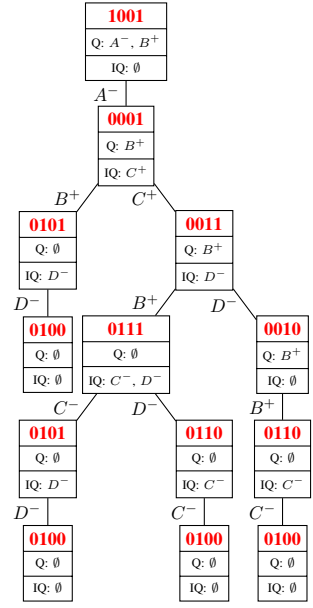

Consider the circuit depicted in \cref{fig:simplecircuit}, which consists of two NOR gates and four wires, $A$, $B$, $C$, and $D$. Herein, $W_I=\{A,B\}$ are circuit inputs, and $W_G=\{C,D\}$ are gate outputs. Assuming an initial state of $(A, B, C, D) = (1, 0, 0, 1)$, \cref{fig:full_tree} shows the complete state-space tree for the input queue $Q=(A^-, B^+)$. Each node in the tree represents a state, and each edge between two nodes represents a transition between two states. Nodes are split into three parts: the first part, highlighted in bold red, describes the state of the circuit as a tuple $(A, B, C, D)$. The second part, prepended with 'Q', is the input queue. The third part, prepended with 'IQ', is the inconsistency queue.

\begin{figure}
    \centering
\subfloat[A simple example circuit without loops]{
\begin{tikzpicture}[ scale=0.78, transform shape, circuit logic US, huge circuit symbols]
    \node [nor gate] (a0) {};
    \draw (a0.input 1) -- ++(-1, 0) node[left] {A};
    \draw (a0.input 2) -- ++(-1, 0) coordinate[pos=0.5] (B_connection) node[left] {B};
    \node [nor gate, anchor=input 1] (a1) at ([xshift=1.5cm]a0.output) {};
    \draw (a0.output) -- (a1.input 1) node[midway, above] {C};
    \draw (B_connection) node[circ]{} -- ++(0, -0.5) -- ++(2, 0) |- (a1.input 2) {};
    \draw (a1.output) -- ++(0.5, 0) node[right] {D};
\end{tikzpicture}
    \label{fig:simplecircuit}}
    \hfil
  \subfloat[A simple circuit with a feedback loop]{
\begin{tikzpicture}[ scale=0.75, transform shape, circuit logic US, huge circuit symbols]
    \node [nor gate] (a0) {};
    \draw (a0.input 1) -- ++(-0.5, 0) node[left] {A};
    \draw (a0.output) -- ++(1, 0) coordinate[pos=0.5] (B_connection) node[right] {B};
    \draw (a0.input 2) -- ++(-0.3, 0) -- ++(0, -0.5) -| (B_connection) node[circ]{};
\end{tikzpicture}
    \label{fig:loop}}
    \caption{\small Two simple example circuits, without and with a feedback loop.}
\end{figure}
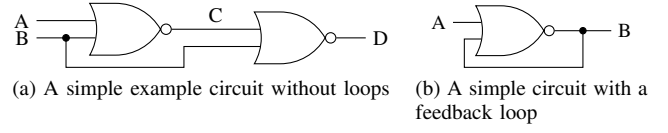

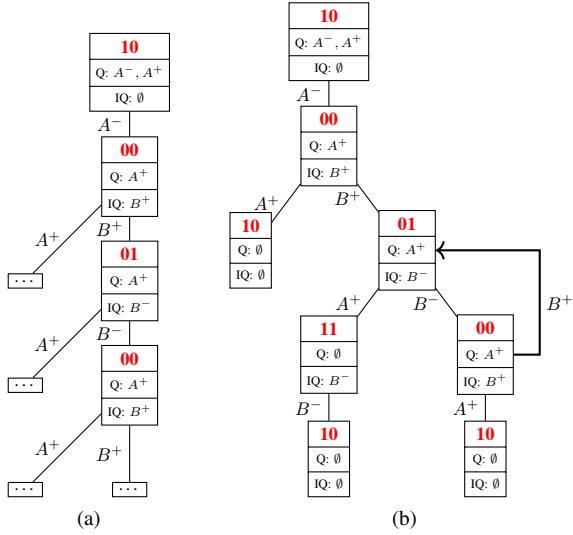
\begin{figure}[h]
  \centering
  \subfloat[]{
  \begin{tikzpicture} [ scale=0.69,       transform shape,
    level distance = 20mm,
    sibling distance = 20mm
]
\node[rectangle split, rectangle split parts=3, draw] {
    \textbf{\color{red}10}
    \nodepart{second} 
    \scriptsize Q: $A^-, A^+$
    \nodepart{third} 
    \scriptsize IQ: $\emptyset$
}
child {
    node[rectangle split, rectangle split parts=3, draw] {
        \textbf{\color{red}00}
        \nodepart{second} 
        \scriptsize Q: $A^+$
        \nodepart{third} 
        \scriptsize IQ: $B^+$
    }
    child {
        node[rectangle, draw] {\ldots}
        edge from parent node[midway, left] {$A^+$}
    }
    child {
        node[rectangle split, rectangle split parts=3, draw] {
            \textbf{\color{red}01} 
            \nodepart{second} 
            \scriptsize Q: $A^+$
            \nodepart{third} 
            \scriptsize IQ: $B^-$
        }
        child {
            node[rectangle, draw] {\ldots}
            edge from parent node[midway, left] {$A^+$}
        }
        child {
            node[rectangle split, rectangle split parts=3, draw] {
                \textbf{\color{red}00}
                \nodepart{second} 
                \scriptsize Q: $A^+$
                \nodepart{third} 
                \scriptsize IQ: $B^+$
            }
            child {
                node[rectangle, draw] {\ldots}
                edge from parent node[midway, left] {$A^+$}
            }
            child {
                node[rectangle, draw] {\ldots}
                edge from parent node[midway, left] {$B^+$}
            }
            child[missing] {}
            edge from parent node[midway, left] {$B^-$}
        }
        child[missing] {}
        edge from parent node[midway, left] {$B^+$}
    }
    child[missing] {}
    edge from parent node[midway, left] {$A^-$}
};
\end{tikzpicture}
    \label{fig:looptree}}   
  \hfil
  \subfloat[]{
\begin{tikzpicture} [ scale=0.69,       transform shape, 
    level distance = 20mm,
    sibling distance = 30mm
]
\node[rectangle split, rectangle split parts=3, draw] {
    \textbf{\color{red}10}
    \nodepart{second} 
    \scriptsize Q: $A^-, A^+$
    \nodepart{third} 
    \scriptsize IQ: $\emptyset$
}
child {
    node[rectangle split, rectangle split parts=3, draw] {
        \textbf{\color{red}00}
        \nodepart{second} 
        \scriptsize Q: $A^+$
        \nodepart{third} 
        \scriptsize IQ: $B^+$
    }
    child {
        node[rectangle split, rectangle split parts=3, draw] {
            \textbf{\color{red}10}
            \nodepart{second} 
            \scriptsize Q: $\emptyset$
            \nodepart{third} 
            \scriptsize IQ: $\emptyset$
        }
        edge from parent node[midway, left] {$A^+$}
    }
    child {
        node[rectangle split, rectangle split parts=3, draw, name=loopparent] {
            \textbf{\color{red}01} 
            \nodepart{second} 
            \scriptsize Q: $A^+$
            \nodepart{third} 
            \scriptsize IQ: $B^-$
        }
        child {
            node[rectangle split, rectangle split parts=3, draw] {
                \textbf{\color{red}11}
                \nodepart{second} 
                \scriptsize Q: $\emptyset$
                \nodepart{third} 
                \scriptsize IQ: $B^-$
            }
            child {
                node[rectangle split, rectangle split parts=3, draw] {
                    \textbf{\color{red}10}
                    \nodepart{second} 
                    \scriptsize Q: $\emptyset$
                    \nodepart{third} 
                    \scriptsize IQ: $\emptyset$
                }
                edge from parent node[midway, left] {$B^-$}
            }
            edge from parent node[midway, left] {$A^+$}
        }
        child {
            node[rectangle split, rectangle split parts=3, draw, name=loopend] {
                \textbf{\color{red}00}
                \nodepart{second} 
                \scriptsize Q: $A^+$
                \nodepart{third} 
                \scriptsize IQ: $B^+$
            }
            child {
                node[rectangle split, rectangle split parts=3, draw] {
                    \textbf{\color{red}10}
                    \nodepart{second} 
                    \scriptsize Q: $\emptyset$
                    \nodepart{third} 
                    \scriptsize IQ: $\emptyset$
                }
                edge from parent node[midway, left] {$A^+$}
            }
            edge from parent node[midway, left] {$B^-$}
        }
        edge from parent node[midway, left] {$B^+$}
    }
    edge from parent node[midway, left] {$A^-$}
};

\draw[->, thick] 
    (loopend.east) -- ++(0.5,0) |- node[pos=0.25, right] {$B^+$} (loopparent.east);
\end{tikzpicture}
    \label{fig:looptreedetection}}    
  \caption{\small Result of the generalized state-space tree construction for \cref{fig:loop}, using the
  duplicate detection \cref{alg:detectcycles}. (a) Simplified symbolic execution with a feedback loop and infinite unrolling. (b) A symbolic execution with a feedback loop and loop detection.}
\end{figure}

Unfortunately, circuits containing feedback loops may lead to infinite state-space trees, i.e., prevent the termination of our construction algorithm. The circuit with a feedback loop depicted in \cref{fig:loop} serves as an example here: starting from the initial state of $(1, 0)$ and an input queue $Q=(A^-, A^+)$, an (infinite) tree sketched in \cref{fig:looptree} would be constructed. To mitigate this issue, we will introduce a compact representation of such loops.

\subsection{Loop detection}

To prevent the infinite unrolling of the state-space tree in the case of feedback loops, we generalize our construction to also allow \emph{cycles} that will be compactly represented by \emph{meta-transitions} (see \cref{sec:metratransitions}). For that purpose, we first identify duplicate states, which are characterized by the following properties: two circuit states are identical if and only if
\begin{itemize}
\item[(i)] their wire states are identical,
\item[(ii)] their remaining input queues are identical,
\item[(iii)] the transition that led into those states is identical. Note that this condition ensures that the first of two paths that successively lead to that state is a prefix of the second, albeit they are obviously not identical.
\end{itemize}

Finding duplicate states during the tree construction is computationally quite cheap, since only direct ancestor nodes need to be considered. The algorithm for detecting and reporting duplicates is shown in \cref{alg:detectcycles}. Obviously, \cref{alg:combine} needs to be adapted for using it for terminating looping behavior; see \cref{sec:integration} for details.

\begin{algorithm}
\caption{\small Procedure detectCycles}
\label{alg:detectcycles}
\begin{algorithmic}
\REQUIRE Node $N$
\ENSURE Detection of a cycle in the tree construction
\STATE $ancestor \leftarrow N.parent$
\WHILE{$ancestor \neq null$}
    \IF{$sameState(N, ancestor)$}
        \RETURN cycle detected
    \ENDIF
    \STATE $ancestor \leftarrow ancestor.parent$
\ENDWHILE
\RETURN no cycle
\end{algorithmic}
\end{algorithm}

Reconsidering the feedback circuit from \cref{fig:loop}, the tree construction starting from the initial state $(A,B)=(1,0)$ and the input queue $Q=(A^-,A^+)$ leads to the generalized state-space tree shown in \cref{fig:looptreedetection}.

Thanks to our duplicate state conditions (i)--(iii) above, the immutability of already constructed nodes of our tree is preserved even in the presence of a loop. After all, the mere existence of a back-edge does not create the need to change the already created nodes representing the loop body. Unfortunately, this will not be the case in the annotated version of our tree described in \cref{sec:delayformulas}, however, since the different paths leading to the loop starting node imply different symbolic timestamps.

\subsection{Complexity of the basic state-space tree construction \label{sec:complexityanalysis}}

Like every symbolic execution approach, our basic tree construction algorithm has a bad worst-case space and time complexity.

\subsubsection{Width of the tree}

The width of the state-space tree, i.e., the out-degree of the nodes, is determined by the number of transitions that can occur in the circuit state $s$ corresponding to a node. Since a state transition in our model is defined by a transition on a single wire, the set of possible next states reachable from $s$ is at most $|W_G|+1$, as all gates could be inconsistent in the state $s$. Only a single circuit input transition is taken from the input queue $Q$ in \cref{alg:enumerate}.

\subsubsection{Depth of the tree}

The major challenge in symbolic execution is managing the combinatorial explosion caused by concurrent transitions. In our setting, concurrent transitions result mainly from inconsistent transitions in the inconsistency queue $N.IC$ of a node $N$, as generated by \cref{alg:enumerate}. The main reason for the high time and space complexity is that the order in which the $k=|N.IC|\leq|W_G|$ transitions can be scheduled is arbitrary, unless information about circuit delays is incorporated (see \cref{sec:delayformulas}): For the basic state-space tree construction, each of the $k!$ permutations must be considered.

Our tree construction hence needs to build the entire subtree rooted in $N$, which contains paths representing all possible sequential orderings of the $k$ transitions, together with newly generated concurrent transitions resulting from successor nodes. The size of this subtree is hence huge: even in the case of an empty input queue $N.Q=\emptyset$, its worst-case size is $\sum_{i=0}^k \frac{k!}{(k-i)!} = k!\sum_{i=0}^k 1/i! < ek!$. In the case of $|N.Q|=m>0$, assuming some worst-case $k_{max}$ for the number of inconsistency transitions generated for \emph{any} node, our tree construction generates a subtree with $m$ layers rooted in $N$, with up to $(k_{max}+1)^i$ nodes in layer $1\leq i \leq m$. Each of the $(k_{max}+1)^m$ nodes at the final layer $m$ generates additional $ek!$ nodes. Summing this up gives a worst-case size of $O\bigl((k_{max}+1)^{m}\cdot k_{max}!\bigr)$ for the entire subtree. Whereas this is clearly a very conservative bound, in particular, because further transitions can invalidate previously generated transitions in some paths, it makes efficient pruning methods mandatory.

\section{Augmentations of the State-Space Tree \label{sec:delayformulas}}

The state-space tree constructed according to \cref{sec:symbolicexecution} fully captures all possible transition orderings that can take place in an execution of a circuit when starting from a given initial
state and a given circuit input queue. However, so far, it does not consider any delay information. In this section, we will augment our basic tree construction by adding symbolic timestamps and related constraints to the nodes, which effectively keep track of the timing of the path leading to a node. Needless to say, our augmentation crucially relies on closed-form analytic delay models such as \cite{ferdowsi2025faithful,FFSS26:Integration,ferdowsi2025drafting}.

The deterministic delay formulas provided by such models can be rewritten to ``absolute'' occurrence time formulas of the form $\tau_G: P \times H \rightarrow \mathbb{R}$ for a given gate $G$, which provide the time of the next output transition based on the following symbolic variables:
\begin{itemize}
  \item \emph{Gate parameters} $P$, representing physical properties affecting the gate delay, like load capacitances and resistances.
  \item \emph{Transition history} $H$, representing previous input and output transitions of the gate, expressed via symbolic timestamps.
\end{itemize}
Indeed, thanks to $H$, it is easy to compute the absolute next output transition time from the relative delay formula provided by the delay model.

Such history-dependent occurrence time formulas are ideally suited for our approach, since the transition history $H$ is explicitly maintained in the path in the state-space tree that leads to a node reached via some output transition of $G$. As a consequence, as described in \cite{TEFS25:arxiv}, we can use a computer algebra system to accumulate the individual relative gate delays occurring throughout a path \emph{without} resorting to the non-differentiable $\min/\max$ operators used in traditional static and dynamic timing analysis approaches. Essentially, this is the reason
why we claimed that the delay models \cite{ferdowsi2025faithful,FFSS26:Integration,ferdowsi2025drafting} effectively
unlocked our novel symbolic timing analysis approach.

\subsection{Tree node symbolic timestamp annotations}
\label{sec:TSannotation}

All that needs to be done to annotate a tree node $N$ created in \cref{alg:enumerate} with a symbolic timestamp $N.TS$ is the following: 
\begin{enumerate}
\item[(a)] If $N$ is reached from its parent node $N'$ via an inconsistency transition $inconsistencyTransition \in N'.IQ$, i.e., a transition caused by the output of some gate $G$, we store the symbolic expression $\tau_G(P,H)$ in $N.TS$, where $H$ is the transition history of $G$ available immediately before the transition to $N$, as extracted from the path leading to $N'$.

\item[(b)] If $N$ is reached via an input transition $inputTransition \in N'.Q$, then $N.TS$ is the symbolic occurrence time of that input transition. The root node does not correspond to a transition and is therefore annotated with the starting time $N.TS=0$.
\end{enumerate}

\subsection{Constraint propagation}

The main utility of the tree node annotation described above is that it enables a very effective pruning method. Recall that our basic state-space tree represents all the different transition orders. When a node branches into multiple successors, each of those represents a \emph{different} execution, in which one specific transition has occurred before the other alternative ones. For example, consider the case where $N.IQ$ contains two alternative inconsistency transitions, say, $A^+$ and $B^+$. Then, $N$ will have two children, $N_{A^+}$ and $N_{B^+}$. In any path containing $N_{A^+}$, transition $A^+$ occurs \emph{before} $B^+$, whereas in any path containing $N_{B^+}$, transition $B^+$ occurs \emph{before} $A^+$.

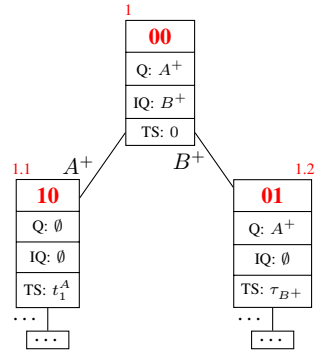
\begin{figure}
\centering
\begin{tikzpicture} [scale=0.85,       transform shape,
    level distance = 25mm,
    sibling distance = 35mm,
    labelcontainer/.style={
        rounded corners=2pt, 
        text=red,
        font=\scriptsize, 
        inner sep=2pt
    }
]
\node[rectangle split, rectangle split parts=4, draw, label={[labelcontainer]110:1}] {
    \textbf{\color{red}00}
    \nodepart{second} 
    \scriptsize Q: $A^+$
    \nodepart{third} 
    \scriptsize IQ: $B^+$
    \nodepart{fourth}
    \scriptsize TS: $0$
}
child {
    node[rectangle split, rectangle split parts=4, draw, label={[labelcontainer]100:1.1}] {
        \textbf{\color{red}10} 
        \nodepart{second} 
        \scriptsize Q: $\emptyset$
        \nodepart{third} 
        \scriptsize IQ: $\emptyset$
        \nodepart{fourth}
        \scriptsize TS: $t_1^A$
    }
    child[level distance=15mm] {
        node[rectangle, draw] {\ldots}
        edge from parent node[midway, left] {\ldots}
    }
    edge from parent node[midway, left] {$A^+$}
}
child {
    node[rectangle split, rectangle split parts=4, draw, label={[labelcontainer]75:1.2}] {
        \textbf{\color{red}01}
        \nodepart{second} 
        \scriptsize Q: $A^+$
        \nodepart{third} 
        \scriptsize IQ: $\emptyset$
        \nodepart{fourth}
        \scriptsize TS: $\tau_{B^+}$
    }
    child[level distance=15mm] {
        node[rectangle, draw] {\ldots}
        edge from parent node[midway, left] {\ldots}
    }
    edge from parent node[midway, left] {$B^+$}
}; 
\end{tikzpicture}
\caption{\small Example of a timestamp-annotated state-space tree.}
\label{fig:constraintpropagation}
\end{figure}

We can leverage this property for a powerful pruning technique that we call \emph{constraint propagation}. Consider the subtree depicted in \cref{fig:constraintpropagation}, where we added a straightforward tree node naming scheme (marked in red above the individual nodes) for reference purposes. Assume that node 1 starts off with a timestamp $1.TS=t_1$. If the execution takes the branch $A^+$ (corresponding to an input transition occurring at time $t_1^A$), the timestamp of the corresponding node 1.1 is ${1.1}.TS=t_1^A$. If the execution takes branch $B^+$ (corresponding to an inconsistency transition), the timestamp of the corresponding node 1.2 is ${1.2}.TS=\tau_{B^+}(P,H)$, where $\tau_{B^+}$ denotes the occurrence time of the transition of the gate causing $B^+$ (which will of course depend on its transition history $H$, which is available in the path leading to node 1). 

Assume first that ${1.1}.TS \neq {1.2}.TS$. Since the branch $A^+$ is only taken if ${1.1}.TS < {1.2}.TS$, we can add this constraint as an additional annotation to node ${1.1}$. Moreover, this constraint can be passed on to all children of node ${1.1}$, since the transition that led to this state can only occur if that constraint is actually satisfied. Analogously, the strict constraint ${1.2}.TS < {1.1}.TS$ can be added to node ${1.2}$ and propagated to its children. Each branch in the tree hence adds branch-ordering constraints and passes them on to its children. Obviously, this continuously restricts the feasible region of the symbolic input times and gate parameters along a given path.

One issue that could invalidate mutual exclusion of propagated constraints arises when two different
child nodes $N_1$ and $N_2$ are annotated with symbolic timestamps that (also) admit a parameter 
assignment that results in $N_1.TS = N_2.TS$. In that case, assuming strict conditions 
$N_1.TS < N_2.TS$ for $N_1$ (resp.\ $N_2.TS < N_1.TS$ for $N_2$) would not cover all possible paths, whereas assuming
non-strict constraints $N_1.TS \leq N_2.TS$ (resp.\ $N_2.TS \leq N_1.TS$) would not be mutually exclusive.
A straightforward solution is to augment symbolic timestamps with deterministic child indices and compare pairs $(N_i.TS,i)$ lexicographically. The input-transition child has index $0$, and the inconsistency transitions in $N.IQ$ are ordered by some fixed gate order. Thus, child $i$ is selected before sibling $j$ iff $N_i.TS<N_j.TS$, or $N_i.TS=N_j.TS$ and $i<j$. Equivalently, the constraint for child $i$ against sibling $j$ is $N_i.TS\leq N_j.TS$ if $i<j$, and $N_i.TS<N_j.TS$ if $i>j$. In the example above, the effective constraint for $N_1$ is therefore $N_1.TS\leq N_2.TS$, whereas the one for $N_2$ is $N_2.TS<N_1.TS$.

Overall, the propagation of constraints ensures two powerful properties:
\begin{enumerate}
  \item Each node is annotated with symbolic constraints that characterize the exact parameter space that leads to that specific execution path, which facilitates a precise analysis of, say, the root causes of timing violations using powerful mathematical tools.
  \item When accruing multiple constraints along a path, conflicts between those can occur at some node $N$, leading to conditions such as $t_1 + 2\delta < t_1+\delta$. Since the resulting parameter space is empty, it is physically impossible for the path leading to $N$ to occur, which allows to prune the complete subtree starting from~$N$.
\end{enumerate}
Implementing the pruning in 2) requires a feasibility checker for the accrued constraints. Since the latter consist of inequalities of path-dependent compositions of the analytic delay formulas provided by the gate delay models, which in turn involve both symbolic input transition times and gate parameters, an SMT solver over the reals or a nonlinear numerical solver is required here. Note that this corresponds to the usage of SAT solvers for path pruning in classic symbolic execution, see \cite{BCDD19:survey}.

\subsection{Efficient loop handling via meta-transitions}
\label{sec:metratransitions}

Whereas the approach described so far works just fine for circuits that do not contain feedback loops, it falls apart when encountering a loop: Since loop detection adds a back edge in the generalized state-space tree, multiple paths lead to the same node. Since these different paths cause different transition histories for the gates involved, the symbolic timestamp originally assigned in \cref{sec:TSannotation} is not applicable for paths involving loop iterations. Moreover, additional constraints originating in the loop body must be added to the node reached by the back-edge and propagated to its children. Since both effects violate our principle of immutability of tree nodes, the entire subtree below the starting node becomes invalid and must be suitably reconstructed.

To tackle this issue, we introduce the concept of \emph{meta-transitions}, which represent multiple transitions as a single transition. Reconsider the NOR gate example with feedback shown in \cref{fig:looptreedetection}, and assume a constant relative gate delay $c>0$ for simplicity. Let the input transition $A^-$ occur at time $t_1$. The state reached immediately after $A^-$ has timestamp $t_1$, and every full loop iteration adds two gate delays. Hence a meta-transition that represents $k\in\mathbb{N}_0$ full loop iterations before the next $B^+$ transition reaches node 1.1.1 at ${1.1.1}.TS=t_1+(2k+1)c$, and the following $B^-$ transition occurs at $t_1+(2k+2)c$. 

To incorporate meta-transitions, we only need to augment a successful cycle detection (\cref{alg:detectcycles}): When a back-edge to some loop starting node $N$ is found, (i) the entire subtree of $N$ in its parent node $N'$ is deleted, and (ii) a new child node $\overline{N}$, reached by a suitable meta-transition, is added to $N'.children$ and to the node $queue$ of the BFS tree construction algorithm. The latter will then reconstruct the subtree for $\overline{N}$, assigning the appropriate symbolic timestamps and constraints. When revisited, cycle detection adds an untraversable \emph{meta-backedge} for efficient loop unrolling, preserving the tree structure.

\begin{figure}
\centering
\begin{tikzpicture} [scale=0.7,       
    transform shape,
    level distance = 30mm,
    sibling distance = 35mm,
    labelcontainer/.style={
        rounded corners=2pt, 
        text=red,
        font=\scriptsize, 
        inner sep=2pt
    }
]
\node[rectangle split, rectangle split parts=4, draw, label={[labelcontainer]120:1}] {
    \textbf{\color{red}10}
    \nodepart{second} 
    \scriptsize Q: $A^-, A^+$
    \nodepart{third} 
    \scriptsize IQ: $\emptyset$
    \nodepart{fourth}
    \scriptsize TS: $0$
}
child {
    node[rectangle split, rectangle split parts=4, draw, label={[labelcontainer]105:1.1}] {
        \textbf{\color{red}00}
        \nodepart{second} 
        \scriptsize Q: $A^+$
        \nodepart{third} 
        \scriptsize IQ: $B^+$
        \nodepart{fourth}
        \scriptsize TS: $t_1$
    }
    child {
        node[rectangle split, rectangle split parts=4, draw, label={[labelcontainer]92:1.1.1}] {
            \textbf{\color{red}10}
            \nodepart{second} 
            \scriptsize Q: $\emptyset$
            \nodepart{third} 
            \scriptsize IQ: $\emptyset$
            \nodepart{fourth}
            \scriptsize TS: $t_2$
        }
        edge from parent node[midway, left] {$A^+$}
    }
    child[edge from parent/.style={draw=red}] {
        node[rectangle split, rectangle split parts=4, draw, name=loopparent, label={[labelcontainer]148:1.1.2}] {
            \textbf{\color{red}01} 
            \nodepart{second} 
            \scriptsize Q: $A^+$
            \nodepart{third} 
            \scriptsize IQ: $B^-$
            \nodepart{fourth}
            \scriptsize TS: ${1.1}.TS + k2c = t_1 + (2k+1)c$
        }
        child[edge from parent/.style={draw=black}] {
            node[rectangle split, rectangle split parts=4, draw=black, label={[labelcontainer]90:1.1.2.1}] {
                \textbf{\color{red}11}
                \nodepart{second} 
                \scriptsize Q: $\emptyset$
                \nodepart{third} 
                \scriptsize IQ: $B^-$
                \nodepart{fourth}
                \scriptsize TS: $t_2$
            }
            child {
                node[rectangle split, rectangle split parts=4, draw=black, label={[labelcontainer]130:1.1.2.1.1}] {
                    \textbf{\color{red}10}
                    \nodepart{second} 
                    \scriptsize Q: $\emptyset$
                    \nodepart{third} 
                    \scriptsize IQ: $\emptyset$
                    \nodepart{fourth}
                    \scriptsize TS: $1.1.2.1.TS + c = t_2 + c$
                }
                edge from parent node[midway, left] {$B^-$}
            }
            edge from parent node[midway, left] {$A^+$}
        }
        child[edge from parent/.style={draw=red}] {
            node[rectangle split, rectangle split parts=4, draw=black, name=loopend, label={[labelcontainer]145:1.1.2.2}] {
                \textbf{\color{red}00}
                \nodepart{second} 
                \scriptsize Q: $A^+$
                \nodepart{third} 
                \scriptsize IQ: $B^+$
                \nodepart{fourth}
                \scriptsize TS: ${1.1.1}.TS + c = t_1 + (2k+2)c$
            }
            child[edge from parent/.style={draw=black}] {
                node[rectangle split, rectangle split parts=4, draw=black, label={[labelcontainer]130:1.1.2.2.1}] {
                    \textbf{\color{red}00}
                    \nodepart{second} 
                    \scriptsize Q: $\emptyset$
                    \nodepart{third}
                    \scriptsize IQ: $\emptyset$
                    \nodepart{fourth}
                    \scriptsize TS: $t_2$
                }
                edge from parent node[midway, left] {$A^+$}
            }
            edge from parent node[midway, left] {$B^-$}
        }
        edge from parent node[midway, left] {$B^+$}
    }
    edge from parent node[midway, left] {$A^-$}
};
\draw[->, color=red] 
    (loopend.east) -- ++(0.5,0) |- node[pos=0.25, right, text=black] {$B^+$} (loopparent.east);

\end{tikzpicture}
\caption{\small A meta-transition representing the loop in \cref{fig:looptreedetection}.}
\label{fig:simpleloopmetatransition}
\end{figure}
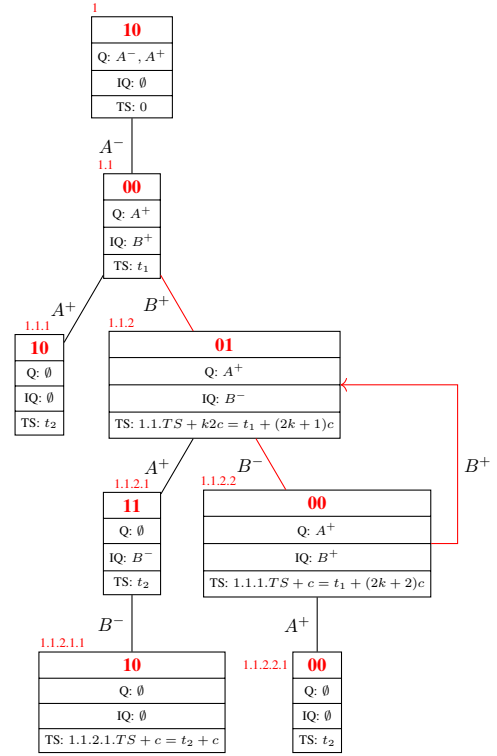

Unfortunately, however, this construction only works for \emph{simple loops}, where the loop body is restricted to a \emph{chain} of nodes reached via a sequence of inconsistency transitions; any additional transition happening in the state corresponding to such a node must exit the loop. Interestingly, simple loops may also be partially overlapping, in the sense that a transition exiting a loop may also start a new loop. Note that the latter may even have a back edge to a node within the loop body of the first loop, provided this node is not the starting node of the first loop.

Circuits encountered in practice may also contain \emph{complex loops}, however: 1) A loop that is properly nested within another one. 2) A loop that has branches in the loop body, which lead to multiple back edges ending in its starting node. It is well-known from classic symbolic execution that dealing with such complex loops is challenging; see, e.g., \cite{XCLL16}. This is, of course, also the case in our specific context, where we face the additional problem that classic solutions such as loop summarization would require us to also incorporate classic symbolic execution for determining the required semantic information. We must, hence, leave a proper handling of complex loops to future research and solely rely on the unrolling of complex loops by iteratively refining the corresponding meta-transitions in our current implementation.

\subsection{Goal functions \label{sec:goalfunction}}

Among the attractive additional features of our symbolic timing analysis approach is the ability to define \emph{analysis goals}. For example, a goal in circuits containing flip-flops could be analyzing the cause of a setup time violation; recall \cref{fig:flipfop_setuphold,fig:setup_hold}. To accomplish this for a given (say, the $k$-th) rising transition of the clock wire $c_k^+$, one needs the symbolic timestamp of two nodes in all paths in the state-space tree: the node $N_{c_k^+}$ reached by $c_k^+$ and the symbolic timestamp of the node $N_{d_\ell^{\pm}}$ reached by the last preceding transition $d_\ell^+$ or $d_\ell^-$. Then, one must analyze the time difference expression $\Delta=N_{c_k^+}.TS - N_{d_\ell^{\pm}}.TS$. For the gate parameterization causing a setup violation, one would observe that $\Delta$ is smaller than the allowed setup time. One could then look at the gradient of $\Delta$ w.r.t.\ the gate parameters to find an alternative gate parameterization that avoids this~violation.

Evidently, for the above example of an analysis goal, it would suffice to generate a state-space tree that only contains $k$ rising clock transitions in any of its paths. Our framework, therefore, provides \emph{goal functions}, which serve a double purpose: 1) Terminating the state-space tree construction when all relevant transitions for some analysis goal have happened, in any path, and 2) Guiding the process of fulfilling the analysis goal after tree construction, ideally (but not necessarily so) in an automated fashion.

For any wire $w \in W=W_G \cup W_I$, $s \in \{+,-\}$, integer $k\geq 1$, and any path $\pi$ in the state-space tree of a circuit, let $w_{k,\pi}^s$ denote the $k$-th transition of type $w^s$ in $\pi$, and $TS(w_{k,\pi}^s)=N_{w_{k,\pi}^s}.TS$ be its symbolic timestamp.  Let $T$ be the set of all such transitions (which also comprise the circuit input transitions, of course), and $TS(T)=\{TS(w_{k,\pi}^s) \mid w_{k,\pi}^s \in T\}$ be the set of all symbolic timestamps. A goal function $g:2^T \to \mathbb{R}^n$, for some finite integer $n>0$, is an arbitrary function of the symbolic timestamps of a finite number of transitions. Note that, for any $S \subseteq T$, $g(S)$ only involves symbolic input transition times and gate parameters of gates determined by $S$.

Given $g(S)$ and $S$, the state-space tree construction can be terminated if all transitions in $S$ have occurred in any path. Subsequently, the analysis goal can be approached by studying $g(S)$ in the constructed tree.

\section{Integrated Symbolic Timing Analysis Framework}
\label{sec:integration}

In this section, we will briefly describe how the basic state-space tree construction algorithm of \cref{sec:constructingtree} is extended to the augmentations described in \cref{sec:delayformulas}, i.e., symbolic timestamp assignment, meta-transitions, constraint propagation, and goal functions. This primarily affects \cref{alg:combine} and \cref{alg:enumerate}; the original node construction \cref{alg:construct} only needs to be augmented by the new components stored in a node, as shown in \cref{alg:constructwithconstraints}.

\begin{algorithm}[h]
\caption{\small Procedure constructNode with augmentations}
\label{alg:constructwithconstraints}
\begin{algorithmic}
\REQUIRE Circuit state $SC$, input queue $Q$, parent node $N'$ ($\emptyset$ for initial node)
\ENSURE A valid tree node $N$
\STATE $N \leftarrow$ newNode()
\STATE $N.SC \leftarrow SC$
\STATE $N.Q \leftarrow Q$
\STATE $N.IQ \leftarrow$ calculateInconsistencies($SC$)
\STATE $N.children \leftarrow \emptyset$
\STATE $N.parent \leftarrow N'$
\IF{$N'=\emptyset$}
\STATE $N.TS \gets 0$
\ELSE
\STATE $N.TS \gets \emptyset$
\ENDIF
\STATE $N.constraints \leftarrow \emptyset$
\RETURN $N$
\end{algorithmic}
\end{algorithm}

The tree construction main \cref{alg:combine} must be extended to identify and handle loops via meta-transitions and to prematurely terminate the further construction of a subtree once the necessary information for the goal function is already available in the current path. \cref{alg:combinewithconstraints} gives the augmented algorithm.

It relies on a procedure detectSimpleCycle($child$), which detects whether the node $child$ is the last node of the loop body of a simple loop. This would cause a back edge to some previous node in the current path that has not been reached by a meta-transition in the tree construction earlier. Our procedure just needs to parse the path leading to $child$ upwards and compare the corresponding states to the state of $child$. If a new $cycle$ is found, the entire subtree of the starting node of $cycle$ is deleted. To also reflect this deletion in the tree construction node $queue$, procedure deleteOrphanedNodes($queue$) is called to delete all now orphaned nodes, and the neighbor construction FOR loop is terminated prematurely. Before termination, a new starting node is generated by the procedure createMetaTransition($cycle$); it is now reached from its parent by a meta-transition and added to the tree construction node $queue$. Finally, the procedure pathIncomplete($child, S$) is used for checking whether the transitions in $S$ needed for the goal function $g(S)$ are already present in the path leading to $child$, by parsing it upwards and collecting the transitions involved.

\begin{algorithm}[h]
\caption{\small Construct tree with augmentations}
\label{alg:combinewithconstraints}
\begin{algorithmic}
\REQUIRE Circuit $C$, initial state $s_0$, initial input queue $Q$, $S$ for goal function $g(S)$
\ENSURE Desired tree constructed
\STATE $rootNode \gets constructNode(s_0,Q,\emptyset)$
\STATE $\begin{aligned}
rootNode.constraints.{}&\mathrm{addAll}(\\
&\mathrm{inputOrderConstraints}(Q))
\end{aligned}$
\STATE $queue \leftarrow \emptyset$
\STATE $queue.$append($rootNode$)
\WHILE{$queue.$length $> 0$}
    \STATE $nextItem \leftarrow queue.$popleft()
    \STATE createChildren($nextItem.SC$, $nextItem.Q$, $nextItem$)
    \FORALL{$child$ in $nextItem.children$}
        \STATE $cycle \leftarrow$ detectSimpleCycle($child$)
        \IF{$cycle \neq \emptyset$}
            \STATE $loopStart \leftarrow cycle.startNode$
            \IF{$loopStart$ is $createdByMetaTransition$}
                \STATE createMetaBackedge($cycle.endNode$, $cycle.startNode$)
            \ELSE
                \STATE $loopStartParent \leftarrow loopStart.parent$
                \STATE $loopStartParent.$delete($loopStart$)
                \STATE deleteOrphanedNodes($queue$)
                \STATE $child \leftarrow$ createMetaTransition($cycle$)
                \STATE $queue$.append($child$)
                \STATE break FOR loop
            \ENDIF
        \ELSE
            \IF{pathIncomplete($child, S$)}
                \STATE $queue$.append($child$)
            \ENDIF    
        \ENDIF
    \ENDFOR
\ENDWHILE
\end{algorithmic}
\end{algorithm}

The creation of all augmented child states is also just an extension of the original algorithm \cref{alg:enumerate}, where node timestamps and constraints are added. \cref{alg:enumeratewithconstraints} uses the procedure computeTimestamp($child, inconsistencyTransition$) to compute the symbolic timestamp for node $child$ according to \cref{sec:TSannotation}.(a), where the gate $G$ (resp.\ the transition history $H$) is extracted from $inconsistencyTransition$ (resp.\ from the path leading to $child$). In addition, every $child$ inherits the parent's constraint set $N.constraints$, which is augmented by the additional sibling-ordering inequalities; the latter express the fact that the transition leading to $child$ precedes every transition leading to its siblings. A child node with accrued constraints that are found to be unsatisfiable is deleted.

\begin{algorithm}[h]
\caption{\small Procedure createChildren with augmentations}
\label{alg:enumeratewithconstraints}
\begin{algorithmic}
\REQUIRE Circuit state $SC$, input queue $Q$, tree node $N$
\ENSURE All valid children added to $N$
\IF{$Q.\mathrm{length} > 0$}
    \STATE $newQueue \leftarrow Q.\mathrm{copy}()$
    \STATE $inputTransition \leftarrow newQueue.\mathrm{popleft}()$
    \STATE $newCircuitstate \leftarrow SC.\mathrm{copy}()$
    \STATE $newCircuitstate.\mathrm{applyTransition}(inputTransition)$
    \STATE $\begin{aligned}
child \leftarrow {}&\mathrm{constructNode}(\\
&newCircuitstate,newQueue,N)
\end{aligned}$
    \STATE $child.TS \leftarrow inputTransition.time$
    \STATE $N.\mathrm{addChild}(child)$
\ENDIF
\FORALL{$inconsistencyTransition$ in $N.IQ$}
    \STATE $newCircuitstate \leftarrow SC.\mathrm{copy}()$
    \STATE $\begin{aligned}
newCircuitstate.{}&\mathrm{applyTransition}(\\
&inconsistencyTransition)
\end{aligned}$
    \STATE $\begin{aligned}
child \leftarrow {}&\mathrm{constructNode}(\\
&newCircuitstate,Q,N)
\end{aligned}$
    \STATE $\begin{aligned}
child.TS \leftarrow {}&\mathrm{computeTimestamp}(\\
&N,inconsistencyTransition)
\end{aligned}$
    \STATE $N.\mathrm{addChild}(child)$
\ENDFOR
\FORALL{$child$ in $N.children$}
    \STATE $child.constraints.addAll(N.constraints)$
    \STATE $C_s \leftarrow \mathrm{computeSiblingConstraints}(child,N.children)$
    \STATE $child.constraints.addAll(C_s)$
\ENDFOR
\FORALL{$child$ in $N.children$}
    \IF{$\mathrm{unsatisfiable}(child.constraints)$}
        \STATE $N.children.remove(child)$
    \ENDIF
\ENDFOR
\end{algorithmic}
\end{algorithm}

\subsection{Timing analysis based on the augmented state-space tree \label{sec:extractingparameters}}

Once the augmented state-space tree is fully constructed, it can be used for various timing analysis purposes. After all, unlike traditional static and dynamic timing analysis approaches, which generate only corner-case results or numerical results of a single simulated execution, our tree stores every possible transition ordering, for any choice of input transition times and gate parameters. Since the purpose of this paper is to describe the cornerstones of the symbolic execution framework underlying our approach, we will mention just two possibilities.

An important problem in circuit validation is detecting hazards, like two or more transitions happening in an unwanted order. Note that avoiding this is particularly important for asynchronous circuit designs based on relative timing constraints \cite{SGR03:TVLSI,SRG99:ASYNC}. A particularly important question here is under which circuit input timing conditions such a hazard could happen. This question can be answered by identifying paths containing such hazards and solving the accrued path constraints. 

In addition, for the particular hazard of a setup time violation, one could use the goal function sketched in \cref{sec:goalfunction} for finding a gate parametrization $P$ that maximizes the setup time, be it in the worst-case path, in some specific paths, or in all paths with a fixed sequence of circuit input transition times. Since each fixed path yields analytic expressions on its feasible constraint region, this can be approached by constrained local optimization methods, such as gradient-based methods applied separately to the relevant path regions.

\subsection{Experimental results \label{sec:results}}

To assess the principal viability of our approach, we developed a research prototype implementation in Python, which will be made publicly available in the future. The prototype was executed on a workstation with an AMD Ryzen 5 3600 6-Core Processor and 8 GB of RAM. For our experiments, we chose the circuit c17\_slack from the ISCAS85 benchmarking library \cite{ISCAS85_reference}, utilizing a simple constant delay model with variable-length circuit input queues. For comparison purposes, we ran our experiments with and without constraint propagation enabled and recorded both the execution time in seconds and the number of candidate paths found. 

\begin{table}
\caption{{\small Prototype results on the c\_17 slack circuit.  $|Q|$ is the input queue size, $Et$ ($s$) is the execution time in seconds, and $\#C$ is the number of paths explored.}}
\centering
\scalebox{0.97}{
\begin{tabular}{l|ll|ll|ll|}
\cline{2-7}
                                         & \multicolumn{2}{l|}{\textbf{No constraint prop.}} & \multicolumn{2}{l|}{\textbf{With constraint prop.}} & \multicolumn{2}{l|}{\textbf{Difference (ratio)}}          \\ \hline
\multicolumn{1}{|l|}{$|Q|$} & \multicolumn{1}{l|}{$Et$ (s)}  & \#C & \multicolumn{1}{l|}{Et (s)}  & \#C & \multicolumn{1}{l|}{Et (s)} & \#C  \\ \hline
\multicolumn{1}{|l|}{1}                  & \multicolumn{1}{l|}{0.095}               & 1                 & \multicolumn{1}{l|}{0.141}               & 1                 & \multicolumn{1}{l|}{\textbf{1.48}}           & \textbf{1}       \\ \hline
\multicolumn{1}{|l|}{2}                  & \multicolumn{1}{l|}{0.093}                & 3                 & \multicolumn{1}{l|}{0.11}                & 3                 & \multicolumn{1}{l|}{\textbf{1.18}}  & \textbf{1}       \\ \hline
\multicolumn{1}{|l|}{4}                  & \multicolumn{1}{l|}{0.16}                & 112               & \multicolumn{1}{l|}{0.218}               & 15                & \multicolumn{1}{l|}{\textbf{1.36}}  & \textbf{0.13}    \\ \hline
\multicolumn{1}{|l|}{8}                  & \multicolumn{1}{l|}{185.93}              & 234708            & \multicolumn{1}{l|}{2.801}               & 153               & \multicolumn{1}{l|}{\textbf{0.015}} & \textbf{0.00065} \\ \hline
\multicolumn{1}{|l|}{16}                 & \multicolumn{1}{l|}{--}                   & --                 & \multicolumn{1}{l|}{27.283}              & 1991              & \multicolumn{1}{l|}{--}              & --                \\ \hline
\end{tabular}}
\label{tab:benchmarks}
\end{table}

Our results, which are summarized in \cref{tab:benchmarks}, show a significant reduction in the number of candidate paths explored when using constraint propagation, especially when the length of the input queue increases. We are convinced that additional complexity reduction techniques, like partial order reduction (see \cref{app:discussion}), in addition to engineering improvements such as a C/C++ implementation and utilizing multi-threading, will allow us to significantly improve the performance and scalability even further.

\section{Conclusions}
\label{sec:conclusions}

In this paper, we introduced a novel symbolic execution framework for symbolic timing analysis of digital integrated circuits, which has been unlocked by our recent advances in accurate gate delay models. Its purpose is to provide an alternative to traditional simulation-based dynamic timing analysis approaches, which do not allow for the complete exploration, e.g., of the root causes of static timing analysis violations.

\balance
\bibliographystyle{IEEEtran}
\bibliography{references}

\clearpage
\nobalance
\appendix
\label{Sec_Appendix}
This appendix sketches our ongoing work on
partial order reduction (POR), which shall help us to reduce the complexity of our
symbolic timing analysis approach even further.

\subsection{Partial Order Reduction}
\label{app:discussion}

Since the main body of our paper is devoted to the core 
elements of our symbolic timing analysis framework,
namely, basic tree construction and its augmentation, we only
presented constraint propagation as a (albeit effective) way for pruning paths
and thus fight the bad worst-case complexity inherent in any symbolic execution 
approach.

Incorporating additional complexity reduction techniques is an important goal of our current work, however, and \emph{partial order reduction} (POR) is among the
most promising candidates.
In fact, the dominant source of the worst-case bound in \cref{sec:complexityanalysis}
is not the (admittedly usually quite large) number of reachable circuit states, but rather the explicit
total ordering of transitions that are enabled at the same node: If $r$ enabled transitions are mutually independent, the current construction may generate up to $r!$ paths that differ only in their order. This situation is frequently encountered in modern circuits, which typically consist of many loosely-coupled subcomponents that can operate almost independently of each other.

 In a nutshell, POR represents such situations by means of \emph{equivalence classes} of finite path segments: each segment is obtained by interleaving transition sequences from different components arbitrarily, while preserving the internal order of every such sequence. This is sound only when transitions belonging to different components commute, i.e., when their relative order does not affect the resulting Boolean state, the remaining input queue, or the symbolic timing information needed later.

The cornerstones of the POR extension of our tree construction that we are about to implement are the following:
\begin{enumerate}
\item[(1)] We distinguish two types of tree nodes, the already introduced \emph{ordinary nodes} and \emph{component nodes}, which belong to some \emph{component} of a \emph{POR region}. Note that both node types have the same internal structure.
\item[(2)] A POR region is created by $k \geq 2$ children of some ordinary node $N$, which is the \emph{starting node} of the POR region; note that $N$ itself is not considered part of its POR region. The $k$ transitions $T_1,\dots,T_k$ leading to the children $N_1,\dots,N_k$ must be pairwise independent, in the (conservative) sense that executing them in different orders neither disables nor modifies each other and preserves the symbolic timing information relevant to subsequent delay computations. (If this independence cannot be established, the ordinary full subtree construction must be resorted to instead). The component node $N_i$ is the \emph{starting node} of the \emph{component} $C_i$ of the POR region of $N$.
\item[(3)] Every component $C_i$ of a POR region of $N$ consists of a subtree of component nodes, constructed analogously to an ordinary subtree but in a restricted mode: A transition is added to $C_i$ only if its enabling and symbolic timestamp can be computed from the common prefix up to $N$ and from transitions already present in $C_i$, and if, in accordance with (2), adding it does affect 
any transition in another component $C_j$.
\item[(4)] If, in the course of the construction of some component's $C_i$'s subtree, some leaf node would (also) be extended by a transition that violates (3), then the POR region of $N$ needs to be refined: All involved components ($C_i$ and at least one $C_j$) are deleted from the POR region of $N$, whereas the other components are not touched. Moreover, the starting nodes of the deleted components ($N_i$ and at least one $N_j$) are re-instantiated as ordinary children of $N$, and the ordinary tree construction is used to build the full subtrees below them.
\item[(5)] POR regions may also be recursive, in the sense that the component node subtree of a component $C_i$ of the POR region of $N$ may contain another POR region started by some of its nodes.  
\end{enumerate}

It is apparent that POR regions are a way to \emph{defer} the full construction of the subtree of some node (i.e., a component $C_i$'s starting node $N_i$) until the construction process has figured out that some ordering of the transition sequences in $C_i$ and in some $C_j$ is inevitable. Before that, these transition sequences can be interleaved arbitrarily. Note carefully that, as soon as the construction process has constructed all paths in the tree up to the worst-case path length (recall \cref{fig:worst_case}) in the circuit, all POR regions are usually gone (unless the circuit has a tree-like structure). Before reaching that path length, however, POR regions may substantially reduce the size of the tree. It is hence particularly interesting in conjunction with goal formulas, as they usually allow for avoiding constructing the whole tree.

We conclude this section by highlighting that POR is complementary to constraint propagation: Constraint propagation allows for pruning paths in the tree as soon as the accrued timing constraints are
physically infeasible. POR just defers the construction of some subtrees until it is inevitable.

\end{document}